\documentstyle[12pt]{article}

\newlength{\dinwidth}
\newlength{\dinmargin}
\def\lapproxeq{\lower .7ex\hbox{$\;\stackrel{\textstyle
<}{\sim}\;$}}
\def\gapproxeq{\lower .7ex\hbox{$\;\stackrel{\textstyle
>}{\sim}\;$}}

\begin{document}
\titlepage
\begin{flushright}
DTP/96/02  \\
February 1996 \\
\end{flushright}

\begin{center}
\vspace*{2cm}
{\Large \bf Constraints on gluon evolution at 
small $x$\footnote{Submitted to Zeitschrift f\"ur Physik {\bf C}}
} \\
\vspace*{1cm}
J.\ Kwieci\'{n}ski\footnote{On leave from Henryk
Niewodnicza\'{n}ski Institute of Nuclear Physics, 31-342
Krak\'{o}w, Poland.}, A.\ D.\ Martin and P.\ J.\ Sutton, \\
\vspace*{0.5cm}
Department of Physics, University of Durham, Durham, DH1 3LE,
England
\end{center}

\vspace*{5cm}
\begin{abstract}
The BFKL and the unified angular-ordered equations are solved to
determine the gluon distribution at small $x$.  The impact of
kinematic constraints is investigated.  Predictions are made for
observables sensitive to the gluon at small $x$.  In particular
comparison is made with measurements at the HERA electron-proton
collider of the proton structure function $F_2 (x, Q^2)$ as a
function of $\ln Q^2$, the charm component, $F_2^c(x,Q^2)$ and 
diffractive $J/\psi$ photoproduction.
\end{abstract}

\newpage
\noindent {\large \bf 1.  Introduction}

Understanding the small $x$ behaviour of the gluon
density is one of the most challenging problems of perturbative
QCD.  It has become topical with the commissioning of the
electron-proton collider HERA at DESY.  Indeed experiments at
HERA are probing the gluon density in the region $x \lapproxeq
 10^{- 3}$ by
observation of the behaviour of the proton structure
function\footnote
{In deep inelastic scattering we do not probe the gluon directly,
but instead via the process $\gamma g \rightarrow q
\overline{q}$. The
longitudinal
fraction $x$ of the proton's momentum that is carried by the
gluon is
therefore sampled over an interval bounded below by the Bjorken
$x$ variable
 $x_{b} \equiv 
Q^2/2p.q$, where as usual $p$ and $q$ are the four momenta of the
incoming 
proton and virtual photon respectively and $Q^2 \equiv -q^2$. We
omit the subscript $b$ since the distinction between $x$ (of the
gluon 
distribution) and $x_{b}$ in $F_2(x_{b},Q^2)$ is evident.} 
$F_2$ as a function of $\ln Q^2$ and by the measurements of
diffractive $J/\psi$ photoproduction, dijet production etc.
\cite{h1,zeus,jpsi}.

First we recall that for moderate values of $x$, say $x
\gapproxeq 0.05$, observable quantities such as $F_2 (x, Q^2)$
are determined in perturbative QCD by the mass factorization
theorem in which the collinear logarithmic singularities, that
arise from gluon emissions in the partonic subprocesses, are
absorbed into universal parton densities.  The absorption of the
collinear singularities make the densities \lq\lq run" with a
$Q^2$ dependence determined by the Altarelli-Parisi (GLAP)
evolution equations, although the absolute values of the
densities are not calculable in perturbative QCD but have to be
input at some scale, say $Q_0^2$.  In fact Altarelli-Parisi
evolution resums the leading $\alpha_S \ln (Q^2/Q_0^2)$
contributions where, in a physical gauge, the $\alpha_S^n \ln^n
(Q^2/Q_0^2)$ contribution is associated with a space-like chain
of $n$ gluon emissions in which the successive gluon transverse
momenta are strongly ordered along the chain, that is $q_{T1}^2
\ll \ldots \ll q_{Tn}^2 \ll Q^2$.  The coefficient (splitting)
functions and anomalous dimensions of Altarelli-Parisi evolution
have been calculated to next-to-leading order.  This corresponds
to the situation in which a pair of gluons are emitted without
strong $q_T$ ordering (and iterations of this configuration). 
Then the contributions contain a power of $\alpha_S$ 
unaccompanied by $\ln(Q^2/Q_0^2)$.

At sufficiently high electron-proton c.m.\ energy, $\sqrt{s}$, we
encounter a second large variable, $1/x \sim s/Q^2$, and we
must resum the leading $\alpha_S \ln (1/x)$ contributions.  In
this regime the dominant parton is the gluon and the key
ingredients to calculate hard scattering observables are the
$k_T$-factorization theorem and the BFKL equation for the gluon
distribution $F (x, k_T^2)$ unintegrated over its transverse
momentum $k_T$ \cite{ktfac,ciafkt}.  The BFKL equation, which
sums the leading $\alpha_S \ln (1/x)$ contributions, may be
written in the form \cite{bfkl,ciaf}
\begin{equation}
F (x, k_T^2) \; = \; F^{(0)} (x, k_T^2) \: + \:
\overline{\alpha}_S \: \int_x^1 \: \frac{dz}{z} \: \int \:
\frac{d^2 q_T}{\pi q_T^2} \: \left [ F ({\frac{x}{z}},
| \vec{k}_T + \vec{q}_T |^2 ) \: - \: \Theta (k_T^2 - q_T^2) \: F
({\frac{x}{z}}, k_T^2) \right ]
\label{a1}
\end{equation}
\noindent with $\overline{\alpha}_S \equiv 3 \alpha_S/\pi$, and
where $F^{(0)}$ is the non-perturbative driving term of the
equation.  To gain insight into the equation we refer to Fig.\ 1
which shows the basic unit that, on iterating (\ref{a1}),
leads to a chain of sequential gluon emissions.  These real
emissions come from the first term under the integral, where
$\vec{k}_T + \vec{q}_T = \vec{k}_T^\prime$ of Fig.\ 1.  The
second term corresponds to the virtual contributions which 
arise from 
gluon loops along the chain.  The real and virtual
contributions cancel as $q_T \rightarrow 0$ to ensure that there
is no singularity in (\ref{a1}).  The strong ordering in $q_T$ is
now no longer applicable and instead we have a \lq\lq random
walk" or diffusion in $\ln q_T^2$ as we proceed along the chain
to gluons with smaller and smaller values of $x$.  The enlarged
$q_T$ phase space leads to a $x^{- \lambda}$ type growth of $F$
as $x \rightarrow 0$.  In fact, since the kernel of the BFKL
equation is scale invariant, the solution can be found by Mellin
transform techniques and given analytically for $x \rightarrow 0$
(for fixed coupling $\alpha_S$).  In particular we have the
famous result, obtained in the original BFKL papers \cite{bfkl},
that 
\begin{equation}
\lambda \; = \; \overline{\alpha}_S \: 4 \ln 2
\label{a2}
\end{equation}
\noindent corresponding to the maximum of a continuous spectrum
of eigenvalues of the kernel of the (Mellin-transformed) BFKL
equation.

The BFKL equation sums the leading order $\alpha_S \ln (1/x)$
contributions.  The next-to-leading terms are not yet known. 
However, there is one kinematical constraint which should be
implemented, namely \cite{ciaf,bo}
\begin{equation}
k_T^2 \; > \; z q_T^2.
\label{a3}
\end{equation}
\noindent The constraint\footnote{Throughout we call this
a kinematic constraint although its origin is partly dynamical. 
We sketch the derivation of (\ref{a3}) in Section 3.} arises
since, in the small $z$ regime where the BFKL equation is valid,
we require that the virtuality of the exchanged gluons arises
mainly from the transverse, rather than the longitudinal,
components of their momentum; that is
\begin{equation}
| k^\prime |^2 \; \approx \; k_T^{\prime 2}.
\label{a4}
\end{equation}

There are also constraints from energy-momentum conservation
\cite{fs}.  For example, for the gluon of transverse momentum
$k_T^{\prime}$ forming one of the links of the BFKL chain we
require $(q+k^\prime)^2 > 0$, where $q$ is the four momentum of
the deep inelastic photon probe. That is
\begin{equation}
k_T^{\prime 2} \; \lapproxeq \; \frac{Q^2}{x} \; \sim \; W^2
\label{a5}
\end{equation}
\noindent at small Bjorken $x$. However (\ref{a3}) gives an
implicit bound on
$k_T^{\prime 2}$ which is usually much more restrictive than 
(\ref{a5}). This can be seen by considering a given value of
$k_T^2.$
Then a high value of $q_T^2$ implies an equally high value of 
$k_T^{\prime 2}$ and (\ref{a3}) becomes
\begin{equation}
k_T^{\prime 2} \; < \; \frac{k_T^2}{z}.
\label{a6}
\end{equation}
\noindent Since $z > x$, this bound is tighter than (\ref{a5})
except
for low values of $Q^2$ which are much less than $k_T^2$. This 
is fortunate because it is possible to study the effects of
imposing (\ref{a3}) on the solution of the BFKL equation.
Indeed, the
introduction of the upper limit of integration,
$q_T^2 < k_T^2/z$, preserves the scale invariance of the BFKL
equation and therefore allows the Mellin transform technique to
be used to find the modified form of the solution as $x
\rightarrow 0$, see Section 2.  The modifications enter at
next-to-leading order and above.  They are found to reduce the
exponent $\lambda$ of the $x^{- \lambda}$ behaviour from the BFKL
value given in (\ref{a2}). Besides finding the analytic solution,
we also solve the BFKL equation numerically and present results
for the effective value of $\lambda$ at non-zero values of $x$ so
that we can study how the asymptotic $(x \rightarrow 0)$ analytic
result is approached.

If $\alpha_S$ is allowed to run then the integrand in the BFKL
equation is weighted more to the infrared end, and the effects of
the kinematic constraint (\ref{a3}) will become weaker. We
investigate the effect in terms of the unified CCFM evolution
equation \cite{ciaf,ccfm,kms1}, which reduces to the BFKL
equation in the small $x$ limit. The CCFM equation is based on
the coherent radiation of gluons, which leads to an angular
ordering in their emissions \cite{dktm}.  The gluon distribution,
which satisfies the CCFM equation, depends on an additional scale
that is required to specify the maximum angle possible for gluon
emission, which is given essentially by the scale of the probe. 
The gluon distribution therefore becomes a function of three
variables $F(x,k_T^2,Q^2)$.  In particular, angular ordering
introduces a constraint
\begin{equation} 
q_T^2 \; < \; \frac{Q^2}{z^2} (1-z)^2
\label{a7}
\end{equation}
\noindent into the integrand of the equation for the gluon. At
small $z$, this angular-ordering constraint $q_T^2 < Q^2/z^2$
appears much weaker than the kinematic constraint $q_T^2 <
k_T^2/z$ of (\ref{a3}). Clearly for $Q^2 \gapproxeq k_T^2$ we
would expect no effect from angular ordering and the gluon
distribution would become independent of $Q^2$. In Section 3 we
solve the CCFM equation numerically, with and without imposing
the kinematic constraint (\ref{a3}), so as to study these
effects.

In Section 4 we use the results for the unintegrated gluon
distribution, together with the $k_T$-factorization theorem, to
study the effect of the kinematic constraint on observable small
$x$ processes. In particular we make predictions for the
behaviour of $F_2(x,Q^2)$ as a function of $\ln Q^2$ at small $x$,
for its charm component $F_2^c(x,Q^2)$, 
and for high energy diffractive $J/\psi$ photoproduction.
Data are available for these three processes, all of which are
sensitive to the gluon at small $x$ \cite{h1,zeus,jpsi}.  We
present our conclusions in Section 5.

\bigskip
\noindent {\large \bf 2.  Constraints on the gluon from the BFKL
equation with fixed $\alpha_S$}

In this section we investigate the effect of imposing the
kinematic constraint $q_T^2 < k_T^2/z$ on the solution of the
BFKL equation, (\ref{a1}).  In particular we wish to solve the
equation for the unintegrated gluon distribution which
incorporates the constraint,
\begin{eqnarray}
F (x, k_T^2) & = & F^{(0)} (x, k_T^2) \nonumber \\
& + & \overline{\alpha}_S \: \int_x^1 \: \frac{dz}{z} \: \int \:
\frac{d^2 q_T}{\pi q_T^2} \; \Theta \left ( \frac{k_T^2}{q_T^2}
\: - \: z \right ) \; \left [ F \left ( \frac{x}{z}, |\vec{k}_T +
\vec{q}_T |^2 \right ) \; - \; \Theta (k_T^2 - q_T^2) \: F \left
(\frac{x}{z}, k_T^2 \right ) \right ], \nonumber \\
& & 
\label{b7}
\end{eqnarray}
\noindent and to determine the value of the exponent $\lambda$ of
the $x^{- \lambda}$ behaviour of the solution.  The solution can
most easily be obtained by rewriting the equation in terms of the
gluon distribution in moment space 
\begin{equation}
\overline{F} (\omega, k_T^2) \; = \; \int_0^1 \; dx \: x^{\omega
- 1} \: F (x, k_T^2).
\label{a8}
\end{equation}
\noindent Then the BFKL equation becomes
\begin{eqnarray}
\overline{F} (\omega, k_T^2) \; = \; \overline{F}^{(0)} (\omega,
k_T^2) & + & \overline{\alpha}_S \: \int \: \frac{d^2 q_T}{\pi
q_T^2} \; \biggl [ \left \{ \int_0^1 \: \frac{dz}{z} \: z^\omega
\; \Theta \left ( \frac{k_T^2}{q_T^2} - z \right ) \right \}
\: \overline{F} (\omega, | \vec{k}_T + \vec{q}_T |^2) \nonumber
\\
& - & \frac{1}{\omega} \; \Theta \; (k_T^2 - q_T^2) \:
\overline{F}
(\omega, k_T^2) \biggr ]
\label{a9}
\end{eqnarray}
\noindent where the first $\Theta$-function imposes the kinematic
constraint on the real emissions.  As usual we solve the equation
by taking the Mellin transform, and introduce the variable $\rho$
conjugate to $k_T^2$
\begin{equation}
\overline{F} (\omega, k_T^2) \; = \; \frac{1}{2 \pi i} \: \int_{c
- i \infty}^{c + i \infty} \: d \rho (k_T^2)^\rho \: \tilde{F}
(\omega, \rho).
\label{a10}
\end{equation}
\noindent then the BFKL equation reduces to the algebraic
relation
\begin{equation}
\tilde{F} (\omega, \rho) \; = \; \tilde{F}^{(0)} (\omega, \rho)
\: + \: \frac{\overline{\alpha}_S}{\omega} \; \tilde{K} (\omega,
\rho) \: \tilde{F} (\omega, \rho)
\label{a11}
\end{equation}
\noindent where the kernel
\begin{equation}
\tilde{K} (\omega, \rho) \; = \; \int \: \frac{d^2 q_T}{\pi
q_T^2} \left [ \left ( \frac{| \vec{k}_T + \vec{q}_T |^2}{k_T^2}
\right )^\rho \: \left \{ \Theta (k_T^2 - q_T^2) \: + \: \left (
\frac{k_T^2}{q_T^2} \right )^\omega \: \Theta (q_T^2 - k_T^2)
\right \} - \Theta (k_T^2 - q_T^2) \right ].
\label{a12}
\end{equation}
\noindent The expression $\{ \ldots \}$ arises from the $\{
\ldots \}$ in
(\ref{a9}), up to a factor of $1/\omega$.  The kernel is scale
invariant, as becomes evident once we make the substitution
$q_T^2 = k_T^2 u$. Then (\ref{a12}) becomes
\begin{eqnarray}
\tilde{K} (\omega, \rho) & = & \int \; \frac{d \phi}{2 \pi} \;
\int_0^1 \; \frac{du}{u} \; \left [ (1 + 2 \sqrt{u} \cos \phi +
u)^\rho \; (1 + u^{\omega - \rho}) \: - \: 1 \right ] \nonumber
\\
& = & \int_0^1 \; \frac{du}{u} \; \left [ F (- \rho, - \rho; 1;
u) \; (1 + u^{\omega - \rho}) \; - \; 1 \right ]
\label{a13}
\end{eqnarray}
\noindent where $F (- \rho, - \rho; 1; u)$ is the hypergeometric
function, and where the $u^{\omega - \rho}$ term arises from the
integration over the interval $1 < u < \infty$ when we substitute
$u \rightarrow 1/u$.

The solution of the BFKL equation for the double Mellin transform
$\tilde{F} (\omega, \rho)$ has the usual form
\begin{equation}
\tilde{F} (\omega, \rho) \; = \; \frac{\tilde{F}^{(0)}(\omega,
\rho)}{1-(\overline{\alpha}_S/\omega) \tilde{K}(\omega, \rho)},
\label{a15}
\end{equation}
\noindent except that now due to presence of the $q_T^2 <
k_T^2/z$ 
cut-off $\tilde{K}$ depends on $\omega$, as well as $\rho$.
Inverting
the moments in the standard way we find that the unintegrated 
distribution is given by
\begin{equation}
F(x,k_T^2) \; = \; \frac{1}{2 \pi i} \int_{c-i \infty}^{c+i
\infty} 
d \rho(k_T^2)^\rho R(\rho)x^{-\omega(\rho)}
\label{a16}
\end{equation}
\noindent where we have performed the $\omega$ integration, and
picked up the \lq\lq leading" pole of $\tilde{F}(\omega, \rho)$
at $\omega = \omega(\rho)$, with residue $R(\rho)$. From
(\ref{a15}) we see that the pole position $\omega (\rho)$ is a
solution of the equation
\begin{equation}
1 - \frac{\overline{\alpha}_S}{\omega(\rho)} \tilde{K}(\omega
(\rho),\rho) \; = \; 0
\label{a17}
\end{equation}
\noindent The exponent $\lambda$ of the $x^{- \lambda}$
behaviour, or in other words the pole position $\overline{\omega}
(\overline{\rho})$ which determines the leading behaviour of
(\ref{a16}) as $x \rightarrow 0$, is the maximum value of
$\omega(\rho)$ as $\rho$
varies along the contour $c-i \infty$ to $c+i \infty$ where $c$
lies in the interval $(-1,\omega)$. The value $\overline{\omega}
(\overline{\rho})$ is found from
\begin{equation}
\frac{d \tilde{K} (\omega(\rho),\rho)}{d \rho} \; = \; 0
\label{a18}
\end{equation}
\noindent which occurs at the saddle point $\rho =
\overline{\rho} =
c$. The resulting
values for the intercept
\begin{equation}
\lambda \; = \; \overline{\omega}(\overline{\rho})
\label{a19}
\end{equation}
\noindent are shown as a function of $\overline{\alpha}_S$ by the
continuous curve in
Fig.2.

The curve is to be compared with the leading-order intercept,
shown as a dashed line, which is obtained from the BFKL equation
without the kinematic constraint imposed. In this case we have
$\omega = 0$ in $\tilde{K} (\omega, \rho)$ in (\ref{a11}) and so
$\tilde{K}$ is a function of $\rho$ alone
\begin{equation}
\tilde{K}(\omega = 0,\rho) \; = \; 2 \Psi(1) - \Psi(1+\rho) -
\Psi(-\rho)
\label{a20}
\end{equation}
\noindent where $\Psi$ is the logarithmic derivative of the Euler
$\Gamma$ function : $\Psi(z) \equiv \Gamma' (z)/\Gamma(z)$. Along
the contour of integration in (\ref{a16}) the pole position
$\omega = \overline{\alpha}_S \tilde{K}$ reaches its maximum
value at the \lq\lq saddle" point $\rho = c = -\frac{1}{2}$, and
so we recover the familiar leading-order BFKL intercept
\begin{equation}
\lambda \; = \; \overline{\alpha}_s \tilde{K} ( \omega = 0, \rho
= -
\: {\textstyle {\frac{1}{2}}}) \; = \; \overline{\alpha}_S 4
\ln 2.
\label{a21}
\end{equation}
\noindent It is easy to see that the imposition of the kinematic
constraint in the BKFL equation introduces higher-order
corrections to the value of the intercept. For instance we note
that the pole in (\ref{a15}) is a solution of
\begin{equation}
\omega \; = \; \overline{\alpha}_S \tilde{K}(\omega = 0, \rho) +
\overline{\alpha}_S \omega  \left . \frac{\partial
\tilde{K}(\omega,
\rho)}{\partial \omega} \right |_{\omega = 0}
+ \ldots. 
\label{a22}
\end{equation}
\noindent If we were to keep only the next-to-leading term in our
determination of the intercept, then we would have
$$
\lambda \; = \; \overline{\alpha}_S \: 4 \ln 2 (1 - 4.15
\overline{\alpha}_S),
$$
\noindent which is shown by the dotted curve denoted by (c) in
Fig.\ 2.  The importance of the higher-order corrections is
evident, even for values of $\overline{\alpha}_S \equiv 3
\alpha_S/\pi \sim 0.1$.

The above analytic procedure yields the \lq\lq asymptotic" or
leading $x^{- \lambda}$ behaviour of the gluon distribution $F
(x, k_T^2)$ as $x \rightarrow 0$.  We may solve the modified BFKL
equation numerically to see how the asymptotic value of the
intercept $\lambda$ is approached as $x$ decreases, for different
values of $k_T^2$.  To be precise we solve (\ref{b7}) numerically
using for the driving term $F^{(0)} (x, k_T^2)$ a conventional
distribution of the form $3 (1-x)^5 N \exp (- k_T^2/k_0^2)$ with
$k_0 = 1$ GeV.  In Fig.\ 3 we show the effective slope
$\lambda_{\rm eff}$, obtained from the solution $F (x, k_T^2)$
via
\begin{equation}
\lambda_{\rm eff} \; = \; \frac{\partial \ln F}{\partial \ln
(1/x)},
\label{b23}
\end{equation}
as a function of $x$ for different values of $k_T^2$, for a
fixed $\overline{\alpha}_S = 0.2$.  We see that the approach to
the asymptotic $(x \rightarrow 0)$ value, $\lambda = 0.35$,
depends on the value of $k_T^2$.  Since $\lambda$ is the leading
singularity we would expect an approach from below.  However, for
large $k_T^2$ the $x$ dependence of the BFKL diffusion pattern,
$$
\exp \; \left ( - A \; \frac{\ln^2 (k_T^2/\overline{k}_T^2)}{\ln
(1/x)} \right ),
$$
\noindent overrides this behaviour and causes $\lambda_{\rm eff}$
to approach the asymptotic value from above.  Here
$\overline{k}_T^2$ is associated with the non-perturbative input
distribution.

\medskip
\noindent {\bf 2.1  Kinematic constraint in the folded form of
the BFKL equation}

In Section 3 we study the effect of the kinematic constraint in
the more realistic case when $\alpha_S$ is allowed to run.  There
we solve the evolution equation in \lq\lq folded" form.  That is
the form in which all the virtual corrections and all the
unresolved real gluon emissions are resummed.  Unresolved
emissions are those with $q_T^2 < \mu^2$, where $\mu^2$ specifies
the resolution.  Imposing the kinematic constraint $q_T^2 <
k_T^2/z$ on the folded BFKL equation is a little more involved
than it was in the \lq\lq unfolded" form (\ref{b7}) in which the
real and virtual terms appear on an equal footing, that is to the
same order in $\alpha_S$.  To see this we first rewrite the
unfolded BFKL equation, (\ref{a9}), in the form 
\begin{eqnarray}
\overline{F} (\omega, k_T^2) \; = \; \overline{F}^{(0)} (\omega,
k_T^2) & + & \frac{\overline{\alpha}_S}{\omega} \int \frac{d^2
q_T}{\pi q_T^2} \: \Theta (k_T^2 - q_T^2) \; \left [ \overline{F}
( \omega, | \vec{k}_T + \vec{q}_T |^2 ) \: - \: \overline{F}
(\omega, k_T^2) \right ] \nonumber \\
& + & \frac{\overline{\alpha}_S}{\omega} \int \frac{d^2 q_T}{\pi
q_T^2} \: \Theta (q_T^2 - k_T^2) \: \left ( \frac{k_T^2}{q_T^2}
\right )^\omega \; \overline{F} (\omega, | \vec{k}_T + \vec{q}_T
|^2 )
\label{b1}
\end{eqnarray}
\noindent where we have simply divided the $q_T$ integration
according to whether $q_T^2$ is less or greater than $k_T^2$, and
carried out the $z$ integrations.  The kinematic constraint is
responsible for the factor ( )$^\omega$ in the second
integral.  Now we divide the first integral according to whether
$q_T^2 < \mu^2$ or $q_T^2 > \mu^2$ so that we can resum all the
virtual and unresolved real terms.  Providing $\mu^2$ is not too
large, we can evaluate the $q_T^2 < \mu^2$ part explicitly:
\begin{eqnarray}
& {\displaystyle \frac{\overline{\alpha}_S}{\omega} \int
\frac{d^2 q_T}{\pi q_T^2}} & \left [\overline{F} (\omega, | k_T +
q_T |^2 ) \; \Theta (\mu^2 - q_T^2) \: - \: \overline{F} (\omega,
k_T^2) \; \Theta (k_T^2 - q_T^2) \right ] \nonumber \\
& & = \; - \: \frac{\overline{\alpha}_S}{\omega} \; F (\omega,
k_T^2) \; \int_{\mu^2}^{k_T^2} \; \frac{dq_T^2}{q_T^2} \: + \: O
\left ( \frac{\mu^2}{k_T^2} \right ) \; \simeq \; - \:
\frac{\overline{\omega}}{\omega} \; F (\omega, k_T^2)
\label{b2}
\end{eqnarray}
\noindent where
\begin{equation}
\overline{\omega} \; \equiv \; \overline{\alpha}_S \ln
(k_T^2/\mu^2).
\label{b3}
\end{equation}
\noindent The result (\ref{b2}) is the residual virtual
correction to $\overline{F} (\omega, k_T^2)$ which remains after
the cancellation of the unresolved real and virtual
singularities.  Using (\ref{b2}), we see that then (\ref{b1})
becomes
\begin{eqnarray}
\overline{F} (\omega, k_T^2) & = & \hat{F}^{(0)} (\omega, k_T^2)
\; + \; \frac{\overline{\alpha}_S}{\omega + \overline{\omega}} \:
\int \: \frac{d^2 q_T}{\pi q_T^2} \; \Theta (k_T^2 - q_T^2) \;
\Theta (q_T^2 - \mu^2) \; \overline{F} (\omega, | \vec{k}_T +
\vec{q}_T |^2 ) \nonumber \\
& & + \frac{\overline{\alpha}_S}{\omega + \overline{\omega}} \;
\int \; \frac{d^2 q_T}{\pi q_T^2} \; \left ( \frac{k_T^2}{q_T^2}
\right )^\omega \; \Theta (q_T^2 - k_T^2) \; \Theta (q_T^2 -
\mu^2) \; \overline{F} (\omega, |\vec{k}_T + \vec{q}_T |^2 ),
\label{b4}
\end{eqnarray}
\noindent where $\hat{F}^{(0)} = \omega \overline{F}^{(0)} =
\omega
\overline{F}^{(0)}/(\omega + \overline{\omega})$.  We invert the
moments and transform back to $x$ space, and find
\begin{eqnarray}
F (x, k_T^2) & = & \hat{F}^{(0)} (x, k_T^2) + \nonumber \\
& & \overline{\alpha}_S \int_x^1 \frac{dz}{z} \int \frac{d^2
q_T}{\pi q_T^2} \; \Theta \left ( \frac{k_T^2}{q_T^2} - z \right
) \; \Theta (q_T^2 - \mu^2) \; \tilde{\Delta}_R (z, k_T^2, q_T^2,
\mu^2) \; F \left ( \frac{x}{z}, | \vec{k}_T + \vec{q}_T |^2
\right ) \nonumber \\
& & 
\label{b25}
\end{eqnarray}
\noindent where $\tilde{\Delta}_R$, often called the non-Sudakov
form factor, is given by
\begin{eqnarray}
\label{c25}
\tilde{\Delta}_R (z, k_T^2, q_T^2, \mu^2) & = & \exp \left \{ -
\overline{\alpha}_S \int_z^1 \frac{dz^\prime}{z^\prime} \; \Theta
\left ( \frac{k_T^2}{q_T^2} - z^\prime \right ) \: \int \:
\frac{dq_T^{\prime 2}}{q_T^{\prime 2}} \; \Theta (k_T^2 -
q_T^{\prime 2}) \; \Theta (q_T^{\prime 2} - \mu^2) \right \}
\nonumber \\
& & \nonumber \\
& & \\
& = & \left \{ \begin{array}{lll}
z^{\overline{\omega}} & {\rm if} & q_T^2 < k_T^2 \nonumber \\ 
& & \\
(z q_T^2/k_T^2)^{\overline{\omega}} & {\rm if} & q_T^2 > k_T^2,
\end{array} \right.
\end{eqnarray}
\noindent where $\overline{\omega}$ is given by (\ref{b3}).  The
form factor screens the $1/z$ singularity in (\ref{b25}).

We see that the inclusion of the kinematic constraint $q_T^2 <
k_T^2/z$ in the real gluon emission part of the unfolded BFKL
equation, (\ref{b7}), has the additional effect of modifying the
non-Sudakov form factor in the folded equation (\ref{b25}).  The
modification occurs for $q_T^2 > k_T^2$ and makes the form factor
larger, but always, of course, satisfying $\tilde{\Delta}_R < 1$.

\medskip
\bigskip
\noindent {\large \bf 3.  Constraints on the gluon from the CCFM
equation} \\

The BFKL equation, which resums the $\alpha_S \ln (1/x)$
contributions, is applicable at small $x$ and moderate $Q^2$,
whereas at larger $x$ and large $(Q^2/Q_0^2)$ Altarelli-Parisi
(GLAP) evolution, which resums the $\alpha_S \ln (Q^2/Q_0^2)$
terms, is appropriate.  A theoretical framework which gives a
unified treatment throughout the $x, Q^2$ region has been
formulated by Ciafaloni, Catani, Fiorani and Marchesini
\cite{ccfm}.  The CCFM approach is based on the coherent
radiation of gluons which implies angular ordering of the gluon
emissions along the chain.  The CCFM equation embodies both the
BFKL and GLAP equations in the appropriate kinematic regimes.  In
the small $x$ region the CCFM equation may be approximated by
\cite{kms1}
\begin{eqnarray}
\label{c1}
F (x, k_T^2, Q^2) & = & F^{(0)} (x, k_T^2, Q^2) \: + \nonumber \\
& & \overline{\alpha}_S (k_T^2) \: \int_x^1 \: \frac{dz}{z} \:
\int \: \frac{d^2 q}{\pi q^2} \; \Theta (Q - zq) \: \Delta_R (z,
k_T^2, q_T^2) \: F \left ( \frac{x}{z}, |\vec{k}_T + \vec{q}_T
|^2, q^2 \right ) \nonumber \\
& & 
\end{eqnarray}
\noindent where it is convenient to impose the angular ordering
in terms of rescaled transverse momenta
\begin{equation}
q \; \equiv \; \frac{q_T}{1 - z}.
\label{c2}
\end{equation}
\noindent The angular-ordering constraint then becomes $q < Q/z$.
Here the non-Sudakov form factor $\Delta_R$ is given by 
\begin{equation}
\Delta_R (z, k_T^2, q_T^2) \; = \; \exp \: \left ( -
\overline{\alpha}_S(k_T^2) \: \int_z^1 \: \frac{dz^\prime}{z^\prime}
\; \int \; \frac{dq^{\prime 2}}{q^{\prime 2}} \; \Theta (k_T^2 -
q^{\prime 2}) \; \Theta (q^\prime - z^\prime q) \right ).
\label{c3}
\end{equation}
\noindent Finally, note that we have allowed $\alpha_S$ to run in
(\ref{c1}).  The solutions $F (x, k_T^2, Q^2)$ of this equation
were studied in ref.\ \cite{kms1}.
As in ref.\ \cite{kms1} we restrict the transverse momenta of
the gluons along the chain to be above 1 GeV$^2$. As we shall 
see it turns out that this physically reasonable choice of
infrared cut-off gives a satisfactory normalization of all the
observables sensitive to the small $x$ behaviour of the gluon
distribution, $F$, with the exception of $F_2^c$ at low $Q^2$.

The novel feature of the CCFM, as compared to the BFKL equation,
is that the solution is dependent on $Q^2$.  The origin of the
dependence comes entirely from angular ordering, $q < Q/z$.  In
the BFKL leading $\ln (1/x)$ limit $F$ becomes independent of
$Q^2$ and indeed the CCFM solutions exhibit this behaviour for
large $Q^2$.  However, for small $Q^2$ non-leading $\ln (1/x)$
effects become important via the angular ordering constraint and
$F$ decreases with decreasing $Q^2$ \cite{kms1}.

Here we study the impact of the kinematic constraint $q_T^2 <
k_T^2/z$.  To be precise the constraint is actually
\begin{equation}
q_T^2 \; < \; (1 - z) \: k_T^2/z.
\label{a33}
\end{equation}
\noindent We sketch the derivation.  The key observation is that
the virtuality $k^2$ of a gluon along the chain should arise
mainly from the transverse, rather than the longitudinal,
components of the momentum for the small $x$ approximation to be
valid.  Now in terms of the light-cone variables $k^\pm \equiv
k_0 \pm k_3$
\begin{equation}
k^2 \; = \; k^+ k^- \: - \: k_T^2
\label{a34}
\end{equation}
\noindent so we require
\begin{equation}
k_T^2 \; > \; | k^+ k^- |.
\label{a35}
\end{equation}
\noindent From Fig.\ 1 we see that
\begin{equation}
k^- \; = \; k^{\prime -} \: - \: q^- \: \simeq \: -q^- \; = \; -
q_T^2/q^+,
\label{a36}
\end{equation}
\noindent where the last equality follows from the on-shell
condition for the emitted gluon\footnote{In (\ref{a36}) and
(\ref{a37}) $q^\pm$ are the light-cone components of the
4-momentum of the emitted gluon, whereas elsewhere in this
section $q$ denotes the rescaled transverse momentum defined by
(\ref{c2}).}.  Thus
\begin{equation}
k^+ k^- \; \simeq \; - \frac{k^+}{q^+} \: q_T^2 \; = \; -
\frac{k^+}{k^{\prime +} - k^+} \: q_T^2 \; = \; - \frac{z}{1 - z}
\: q_T^2.
\label{a37}
\end{equation}
\noindent The kinematic constraint (\ref{a33}) then follows
directly from (\ref{a37}) and (\ref{a35}).

Here we wish to compare the gluon distribution $F (x, k_T^2,
Q^2)$ obtained by solving (\ref{c1}), with the solution obtained
if the kinematic constraint (\ref{a33}) is imposed.  That
is we study the modification of the solution caused by
incorporating 
\begin{equation}
\Theta \: \left ( \frac{k_T^2}{(1 - z) \: q^2} \; - \; z \right )
\quad \hbox{and} \quad \Theta \: \left ( \frac{(1 - z^\prime) \:
k_T^2}{q_T^2} \; - \; z^\prime \right )
\label{a38}
\end{equation}
\noindent in (\ref{c1}) and (\ref{c3}) respectively, just as the
corresponding $\Theta$ functions were included in (\ref{b25}) and
(\ref{c25}).  In the small $z$, large $Q^2$ regime the kinematic
constraint $q_T^2 < (1 - z) k_T^2/z$ is a stronger limitation
than the angular ordering constraint $q^2 < Q^2/z^2$, and we
anticipate that the CCFM solution $F (x, k_T^2, Q^2)$ will become
independent of $Q^2$.  In other words in this limit the kinematic
constraint automatically embodies the angular ordering constraint
\cite{bo} and since the former is independent of $Q^2$ the
unintegrated gluon distribution $F$ does not depend on this
variable either.  However, as $Q^2$ decreases below $k_T^2$ the
angular ordering constraint becomes stronger and $F$ begins to
decrease.  We illustrate the effect in Fig.\ 4 which shows $F$
versus $Q^2$ at different values of $x$ for $k_T^2 = 10 \: {\rm
GeV}^2$.  Here we have calculated $F$ by numerically solving the
CCFM equation (\ref{c1}) as described in ref.\ \cite{kms1}, and
then repeated the calculation with the kinematic constraints
(\ref{a38}) incorporated.  We also use these solutions to show in
Fig.\ 5 the effective exponent $\lambda_{\rm eff}$ of the
integrated gluon distribution
\begin{equation}
\lambda_{\rm eff} \; \equiv \; \frac{\partial \ln (xg)}{\partial
\ln (1/x)} \quad \hbox {with} \quad xg (x, Q^2) \; = \;
\int^{Q^2} F dk_T^2
\label{a39}
\end{equation}
\noindent as a function of $x$ for fixed values of $Q^2$.  The
gluon distributions are generated from a flat input ($F^{(0)}
\sim$ constant as $x \rightarrow 0$), and so the rapid rise with
decreasing $x (xg \sim x^{- \lambda_{\rm eff}})$ is generated by
perturbative QCD resummation effects via the CCFM equation.  As
for the BFKL equation with fixed $\alpha_S$, we notice from Fig.\
5 that for the CCFM equation $\lambda_{\rm eff}$ becomes smaller
(by about 0.1) on imposing the kinematic constraint.

\bigskip
\noindent {\large \bf 4.  Impact on observables}

Predictions for observables at small $x$ are driven by the
behaviour of the gluon distribution, since the gluon is by far
the dominant parton in this regime.  Here we study three
quantities, which are being measured at HERA, that are especially
sensitive to the gluon distribution at small $x$; namely
$\partial F_2/\partial \log Q^2$, the charm component of $F_2$
and high energy diffractive
$J/\psi$ photoproduction.  
The first two observables are linearly dependent on the gluon
density, whereas the third has a quadratic dependence which 
considerably increases its sensitivity to the gluon.

In deep inelastic scattering the virtual photon couples to the
gluon via the $g \rightarrow q\overline{q}$ transition.  We
therefore calculate the structure function $F_2$ from the
unintegrated gluon distribution $F$ using the $k_T$-factorization
theorem
\begin{equation}
F_2 (x, Q^2) \; = \; \sum_q \int \: dk_T^2 \: \int_x^1 \:
\frac{dx^\prime}{x^\prime} \: \int \: d^2 \kappa \: F (x^\prime,
k_T^2, \kappa^2) \: F_q^{\rm box} \left ( \frac{x}{x^\prime},
\vec{\kappa}_T, \vec{k}_T, Q^2,m_q \right ) \: + \: F_2^S
\label{a40}
\end{equation}
\noindent where $F_q^{\rm box}$ includes both the quark \lq\lq box"
and \lq\lq crossed box" contributions which originate from
virtual photon-virtual gluon $q\overline{q}$ production, that 
is from $\gamma g \rightarrow q\overline{q}$.  The convolution is
sketched in Fig.\ 6.  
For the $u,d$ and $s$ quark contributions we take the quark mass
$m_q=0$, while for the charm component we take $m_c=1.5$ GeV.
The explicit expressions for $F_q^{\rm box}$ including quark 
mass effects can be found\footnote{There is a typographical 
error in the expression for $x^{\prime}$ 
below eq.(19) in ref.\ \cite{akms}; 
the factor $\beta (1-\beta)$ should be in the denominator.} 
in ref.\ \cite{akms}; the argument of
$\alpha_S$ in $F_q^{\rm box}$ is taken to be $(\kappa_T^2 + m_0^2)$
where $m_0^2 = 1 $ GeV$^2$ for $u,d,s$ quarks and $m_0^2 = m_c^2$
for the $c$ quark. The results are not very sensitive to 
variations of $m_0$ about these values.

The background contribution $F_2^S \simeq
F_2 (x, Q^2)$ at large $x$, but is a slowly varying function of
$x$ and $Q^2$ at small $x$.  For example we extrapolate below $x
= 0.1$ using the soft Pomeron $x^{- 0.08}$ behaviour.  However,
predictions for the slope, $\partial F_2/\partial \ln Q^2$ are
particularly insensitive to any ambiguities due to $F_2^S$.   
The small $x$ approximation of the CCFM equation, (\ref{c1}),
that we have used amounts to setting the Sudakov form factor
$\Delta_S = 1$ and to approximating the gluon-gluon splitting
function by its singular term as $z \rightarrow 0$, that is
$P_{gg} \simeq 6/z$.  $\Delta_S$ represents the virtual
corrections which cancel the singularities at $z = 1$.  To make a
realistic comparison with data, we allow for the remaining finite
terms in $P_{gg}$ by multiplying the solution $F (x, k_T^2, Q^2)$
by the factor
$$
\exp \; \left ( - \; \frac{33 + 2n_f}{36} \: \int^{Q^2} \:
\overline{\alpha}_S (q^2) \; \frac{dq^2}{q^2} \right )
$$
\noindent where the number of active flavours $n_f = 4$.  At
large $x (x \gapproxeq 0.1)$ and small $Q^2 (Q^2 \lapproxeq 20 \:
{\rm GeV}^2)$ the CCFM solutions, with and without the kinematic
constraint imposed, are found to agree with each other, and also
with the solution obtained from the double-leading-logarithm
(DLL) approximation of (\ref{c1}) in which we replace $\Theta (Q
- zq)$ by $\Theta (Q - q)$ and set $\Delta_R = 1$.  However, at
larger $Q^2$ (beyond the range of the data that we consider here)
some care is needed.  The kinematic constraint is only applicable
in the small $x$ region and so the normalisation of the gluon is
suspect at large $x$, particularly for large $Q^2$.  We therefore
renormalise the solution of the CCFM equation with the kinematic
constraint imposed so as to agree for $x > 0.1$ with the
unmodified solution and its DLL approximation.  In this way we
allow for the small $x$ approximation of the equation.  The
renormalisation only affects the solution for $Q^2 \gapproxeq 20
\: {\rm GeV}^2$.

In Fig.\ 7 we show the predictions for $F_2 (x, Q^2)$ at small
$x$ together with the latest HERA measurements.  Including the
kinematic constraint (\ref{a33}) in the CCFM equation for the
gluon $F (x, k_T^2, Q^2)$ has the effect of taking us from the
dashed to the continuous curves in Fig.\ 7.  The relevant
comparison is the slope $(\partial F_2/\partial \ln Q^2)$ of the
curves which is proportional to the gluon distribution.  With the
present experimental errors the comparison is inconclusive, but
it is evident that, as the statistical and systematic errors are
reduced, future measurements of $\partial F_2/\partial \ln Q^2$
will give insight into the properties of the gluon distribution
$F (x, k_T^2, Q^2)$.

Recently the charm component of $F_2$ has been measured 
\cite{adr} at HERA in the small $x$ region. These measurements
of $F_2^c(x,Q^2)$ are shown in Fig.~8, together with earlier EMC
values \cite{emc} at larger $x$. We compare these data with
the charm component $F_2^c$ determined from the 
{\it unintegrated} gluon distribution, $F$, using the $c$ quark
contribution to the $k_T$-factorization formula (40). We show
the values obtained by taking the mass of the charm quark
to be $m_c=1.4$ and $1.7$ GeV. We also show predictions 
based on GRV \cite{grv} and MRS partons \cite{mrsa}. 
The first of these two is  obtained from $\gamma g \rightarrow 
c \bar{c}$ at NLO \cite{v} using massive charm quarks and 
the {\it integrated} (GRV)
gluon distribution. In the MRS analyses \cite{mrsa,mrsc} the
charm quark is treated as a parton. The charm distribution 
is assumed to be zero for $Q^2 < m^2$, while above this
threshold ($Q^2 > m^2$) it is evolved assuming that 
$m_c=0$. The value $m^2 = 2.7$ GeV$^2$ is determined by 
fitting to the EMC data \cite{emc} for $F_2^c$. 
Although the H1 small $x$ data are 
preliminary, it is clear that
an improved measurement of $F_2^c$ will be valuable.
At present there are indications that our 
$k_T$-factorization approach underestimates the H1 data at the
lower $Q^2$ values; in fact the imposition of the kinematic
constraint worsens our previous description of these
data \cite{kmsp}.

The third observable process that we study is high energy
diffractive
$J/\psi$ photoproduction, $\gamma p \rightarrow J/\psi p$.  It
offers a sensitive probe of the gluon at small $x$
\cite{mrzc,bfg}.  The amplitude can be factored into the product
of the $\gamma \rightarrow c\overline{c}$ transition, the
scattering of the $c\overline{c}$ quark pair on the proton via
(colourless) two-gluon exchange, and finally the formation of the
$J/\psi$ from the outgoing $c\overline{c}$ pair.  The crucial
observation is that at high energy the scattering on the proton
occurs over a much shorter timescale than the $\gamma \rightarrow
c\overline{c}$ fluctuation or the $J/\psi$ formation times.

Since diffractive $J/\psi$ photoproduction is essentially an
elastic process, the cross section is dependent on the {\it
square} of the unintegrated gluon distribution, $F (x, k_T^2,
Q^2)$.  The relevant values of $x$ and $Q^2$ are $x =
M_\psi^2/W^2$ and $\overline{Q}^2 \equiv M_\psi^2/4$, where
$M_\psi$
is the mass of the $J/\psi$ meson and $W$ is the $\gamma p$
centre-of-mass energy.  The cross section is given by \cite{rrml}
\begin{eqnarray}
\label{a41}
\sigma (\gamma p \rightarrow J/\psi p) & = & \frac{1}{b} \: \left
. \frac{d \sigma}{dt} \: (\gamma p \rightarrow J/\psi p) \right
|_0 \nonumber \\
& & \nonumber \\
& = & \frac{\pi^3 M_\psi^3 \alpha_S^2 (\overline{Q}^2)
\Gamma_{ee}}{3 b \alpha} \left | \int \frac{dk_T^2}{k_T^2} \:
\left ( \frac{1}{M_\psi^2} - \frac{1}{M_\psi^2 + 4 k_T^2} \right
) \; F (x, k_T^2, \overline{Q}^2) \right |^2 \nonumber
\\
& & 
\end{eqnarray}
\noindent where $\Gamma_{ee}$ is the leptonic width describing
the $J/\psi \rightarrow e^+ e^-$ decay, $\alpha$ is the QED
coupling, and $b$ is the slope parameter of the differential
cross section, $d\sigma/dt = A \exp (- b |t| )$.  We take the
experimental value $b = 4.5 \: {\rm GeV}^{-2}$.  We include the
effects of $c\overline{c}$ rescattering and the small
contribution of the real part of the amplitude as described in
ref.\ \cite{rrml}.  It was noted in ref.\ \cite{rrml} that the
effects of Fermi motion of the $c$ and $\overline{c}$ quarks in
the $J/\psi$ lead to a sizeable $(\pm 30\%)$ uncertainty in the
normalization of the perturbative QCD prediction of the
photoproduction cross section, but that the \lq\lq shape" of the
$W$ (or $x$) dependence is unaffected.

The predictions for diffractive $J/\psi$ photoproduction are
compared with recent HERA data in Fig.\ 9.  At present the data
extend up to energy $W \simeq 140$ GeV, that is down to $x \simeq
5 \times 10^{- 4}$.  The prediction in the absence of the
kinematic constraint (the dashed curve) implies that the gluon
increases too fast with decreasing $x$.  On the other hand if the
kinematic constraint is incorporated in the CCFM equation, then
the continuous curve is obtained and the description is improved.

For completeness we also show in Fig.\ 9 the description of the
$J/\psi$ data calculated from two recent sets of partons (GRV
\cite{grv} and MRS(A$^\prime$) \cite{mrsa}) as described in ref.\
\cite{rrml}.  Neither parton set incorporates $\ln (1/x)$
resummation effects.  The $J/\psi$ data appear to favour the
phenomenological gluon distribution of the latter set of partons.

\bigskip
\noindent {\large \bf 5.  Conclusions}

We have studied the behaviour of the gluon distribution of the
proton in the small $x$ regime, $10^{-4} \lapproxeq x \lapproxeq
10^{-3}$, that has recently become accessible to the experiments
being performed at the HERA electron-proton collider.  In this
regime it is necessary to work in terms of the unintegrated gluon
distribution $F (x, k_T^2)$ and to resum $\ln (1/x)$
contributions.  To leading order, the resummation is accomplished
by the BFKL equation.  The $\alpha_S^n \ln^n (1/x)$ contribution
corresponds to an effective $n$-rung ladder diagram arising from
a space-like chain of $n$ gluon emissions.  The solution $F (x,
k_T^2)$ of the BFKL equation shows, with decreasing $x$, two
characteristic features.  First an $x^{- \lambda}$ growth (where
$\lambda = \overline{\alpha}_S 4 \ln 2$ for fixed $\alpha_S$, or
$\lambda \simeq 0.5$ if $\alpha_S$ is allowed to run), and
second, a diffusion or random walk in $\ln k_T^2$ as we proceed
along the gluon chain.

The full next-to-leading order summation of terms $\alpha_S^n
\ln^{n - 1} (1/x)$ is not yet known, but we would expect the
description of $F_2$ to be sensitive to this correction.  
However, two important
higher-order effects can 
already
be investigated.  One effect is due to
the angular ordering of gluon emissions, which leads to the CCFM,
rather than the BFKL, equation for the gluon distribution.  The
solution now depends on an additional scale that is required to
specify the maximum angle of gluon emission, which turns out to
be essentially the scale $Q^2$ of the probe.  That the solution
$F (x, k_T^2, Q^2)$ will be dependent on $Q^2$ is indeed 
evident from 
the {\it angular ordering} constraint $\Theta (Q - qz)$ in
the CCFM equation, (\ref{c1}).  At very small $x$ the constraint
is automatically satisfied and the CCFM solution reduces to the
BFKL form $F (x, k_T^2)$.  Numerical solutions to the CCFM
equation were obtained in ref.\ \cite{kms1}.

The second higher-order effect is due to the imposition of the
{\it kinematic} constraint $\Theta (k_T^2/q_T^2 - z)$ which is
required for the validity of the BFKL or CCFM equation at small
$x$.  The constraint is needed to ensure that the virtuality of
the gluons along the chain is controlled by the transverse
momenta, that is $|k^2| \sim k_T^2$.  The
major aim of this paper is to explore the consequences of
implementing this constraint.  For fixed $\alpha_S$ the
introduction of the constraint preserves the scale invariance of
the BFKL equation.  The $x^{- \lambda}$ behaviour of the solution
as $x \rightarrow 0$ can therefore be obtained using analytic
methods.  We found that the BFKL intercept $\lambda =
\overline{\alpha}_S 4 \ln 2$ is significantly reduced.  The
details are shown in Fig.\ 2.  We see that it is insufficient to
consider just next-to-leading effects; higher-order effects are
important.  For running $\alpha_S$ we solved the CCFM equation
numerically and obtained the gluon distribution $F (x, k_T^2,
Q^2)$ with and without the kinematic constraint imposed.  In
this case the constraint has the effect of reducing $\lambda$ by
about 0.1, see Fig.\ 5.  For $Q^2 \gapproxeq k_T^2$ the kinematic
constraint is more severe than that due to angular ordering and
so the gluon distribution becomes independent of $Q^2$, see Fig.\
4.

In Section 4 we studied the impact of imposing the kinematic
constraint on the description of three observables which are
sensitive to the gluon distribution at small $x$ and which are
being measured at HERA.  The observables are $\partial
F_2/\partial \ln Q^2$, $F_2^c(x,Q^2)$ 
and the $W$ dependence of the cross section
for high energy diffractive $J/\psi$ photoproduction.  The
effects are significant in particular regions of phase space;
contrast the dashed and continuous curves in Figs.\ 7 and 9.  As
expected $J/\psi$ photoproduction offers an especially sensitive
measure of the gluon.  
These comparisons with data should be regarded as exploratory,
in other words, our
calculation should be viewed as a preliminary step towards a
full theoretical treatment of higher-order $\ln (1/x)$
contributions.  However, it is clear that the effects discussed
in this paper should be incorporated in any realistic analysis of
improved small $x$ data.

\bigskip
\noindent {\large \bf Acknowledgements}

We thank Dick Roberts for valuable discussions.  J.K.\ thanks the
Department of Physics and Grey College of the University of
Durham for their warm hospitality.  This work has been supported
in part by the UK Particle Physics and Astronomy Research
Council, by Polish KBN Grant No.\ 2 P03B 231 08 and the EU under
Contracts Nos.\ CHRX-CT92-0004 and CHRX-CT93-0357.

\newpage
\noindent {\large \bf Figure Captions}

\begin{itemize}
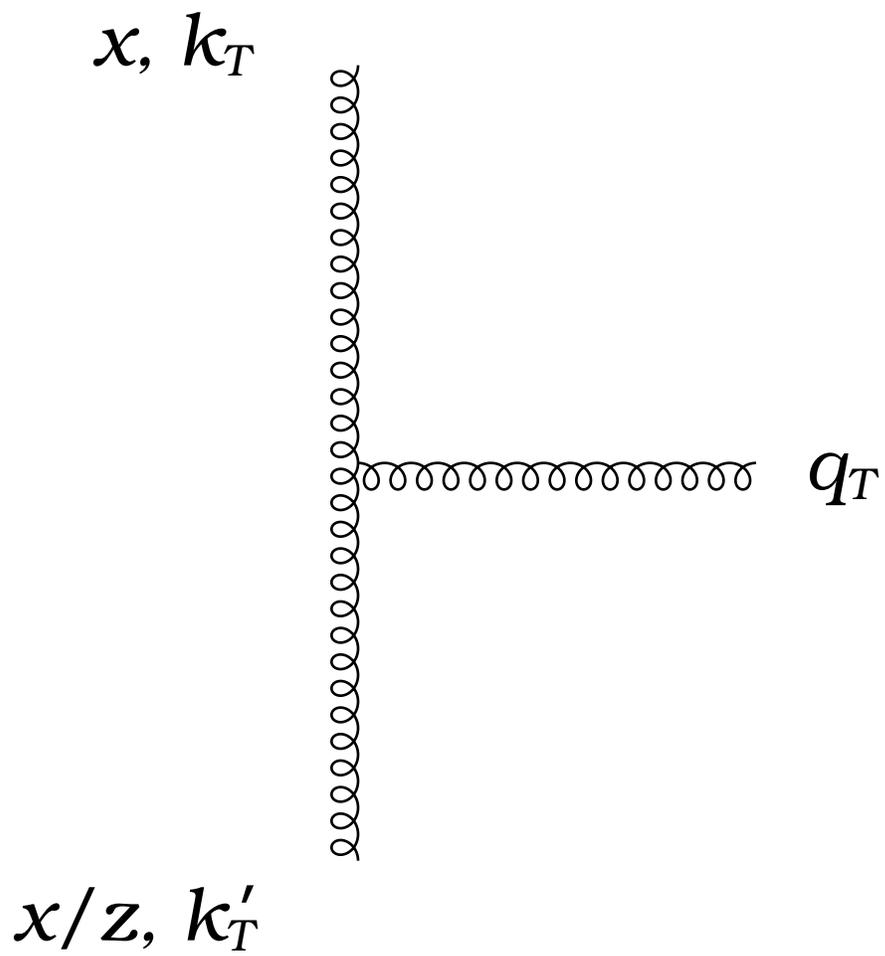
\item[Fig.\ 1] Gluon emission which forms the basis of the BFKL
equation (\ref{a1}) for the unintegrated gluon distribution $F
(x, k_T^2)$.  $x$ and $x/z$ are the longitudinal momentum
fractions of the proton's momentum carried by the respective
gluons.  Throughout we use $q_T$ and $k_T$ to denote,
respectively, the transverse momentum of an emitted gluon and of
a gluon along the chain.

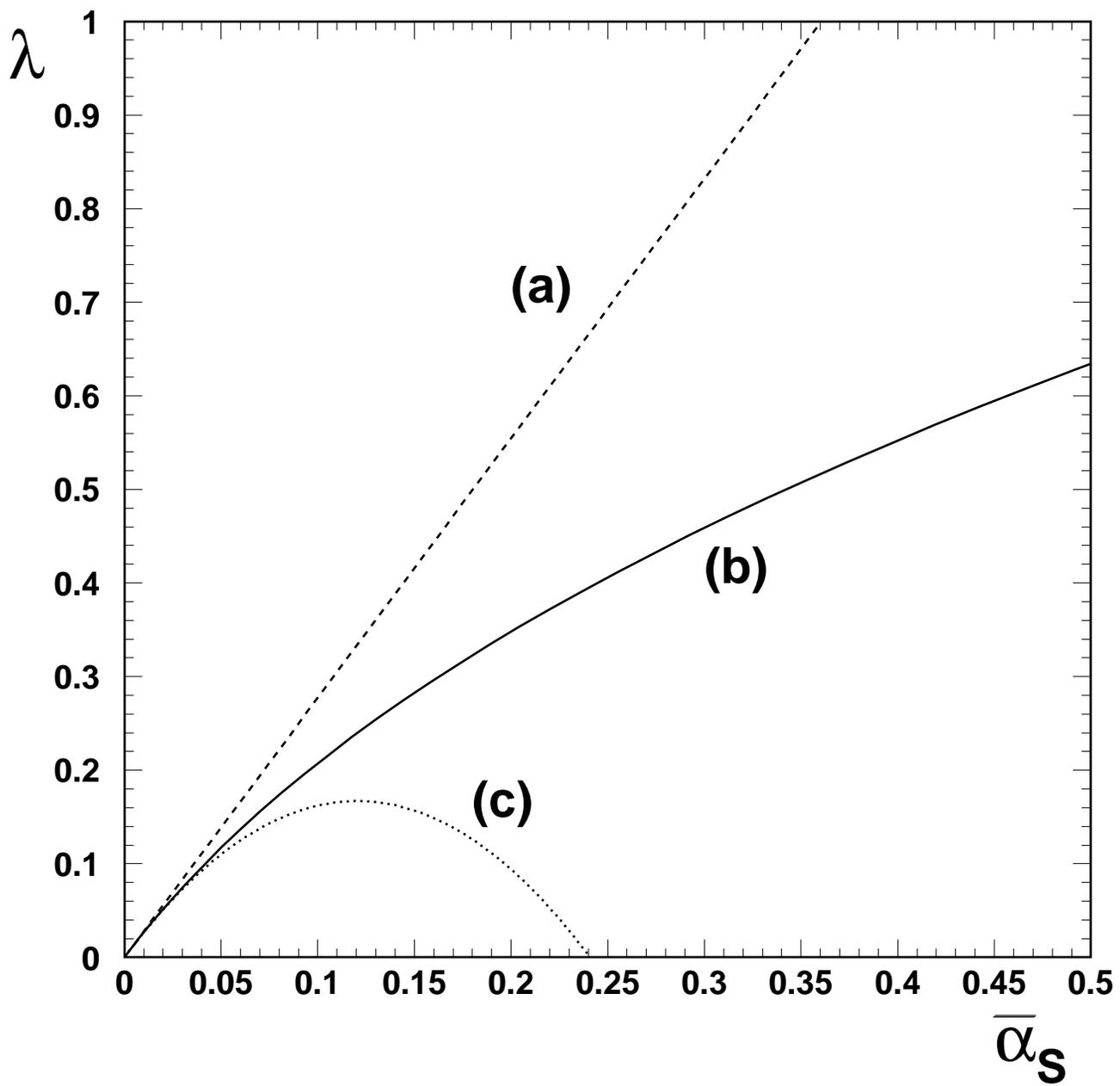
\item[Fig.\ 2] The exponent $\lambda$ of the $x^{- \lambda}$
behaviour of the gluon distribution obtained by solving the BFKL
equation (a) with (continuous curve) and (b) without (dashed
curve) the kinematic constraint imposed, as a function of (fixed)
$\overline{\alpha}_S \equiv 3 \alpha_S/\pi$.  The dashed curve is
$\lambda = \overline{\alpha}_S 4 \ln 2$.  The dotted curve (c) is
the value of the exponent that is obtained if we keep only the
next-to-leading order modification due to the kinematic
constraint.

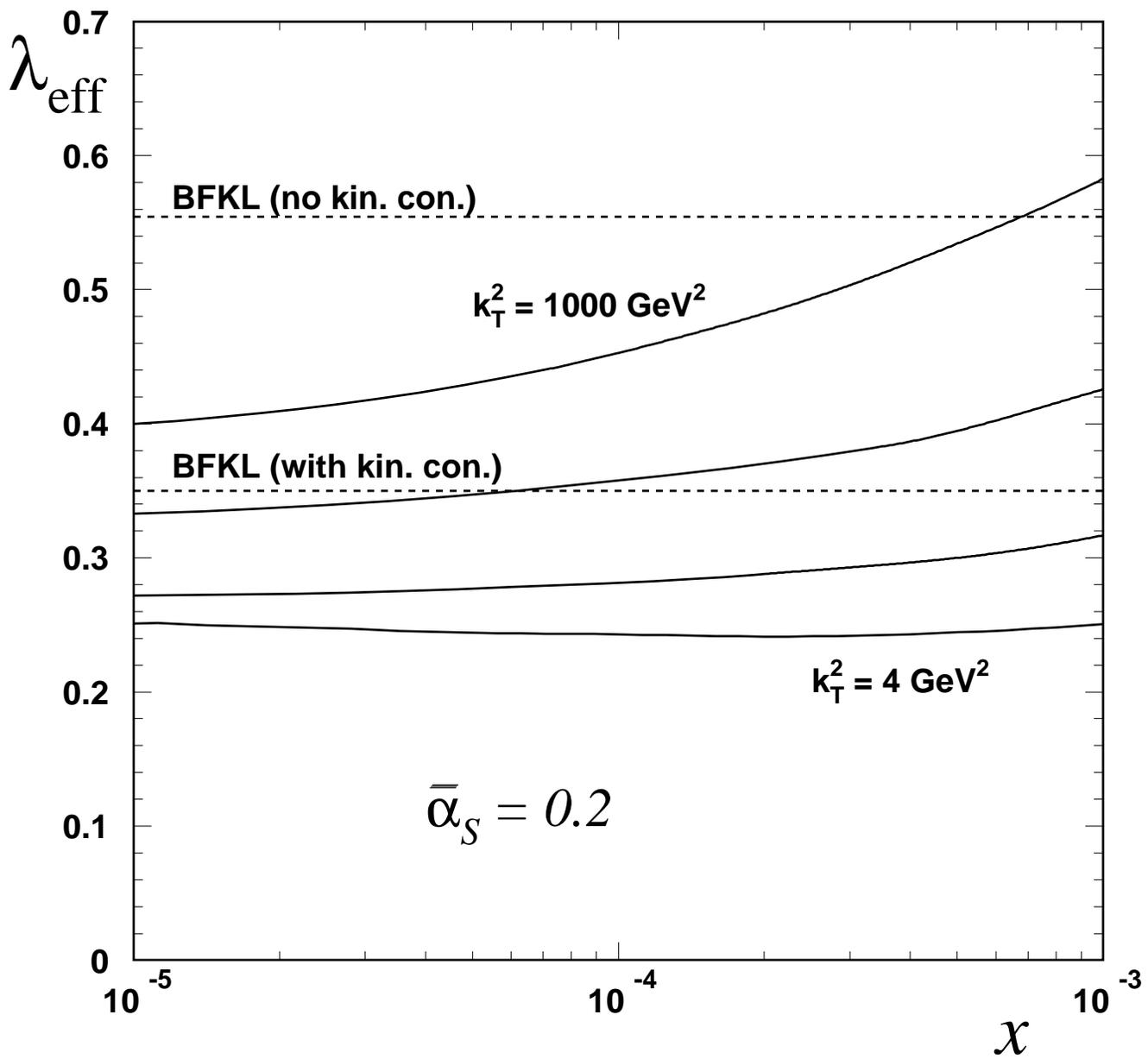
\item[Fig.\ 3] The continuous curves are the effective exponent
$\lambda_{\rm eff} \equiv \partial \ln F/\partial \ln (1/x)$
calculated from the numerical solution $F (x, k_T^2)$ of the BFKL
equation which incorporates the kinematic constraint, for $k_T^2
=$ 1000, 100, 10 and 4 GeV$^2$.  The coupling
$\overline{\alpha}_S \equiv 3 \alpha_S/\pi = 0.2$.  The dashed
lines indicate the value of the exponent of the $x
\rightarrow 0$ analytic solutions of the BFKL equation with and
without the kinematic constraint included, that is the values of
curves (b) and (a) of Fig.\ 2 respectively at
$\overline{\alpha}_S = 0.2$.

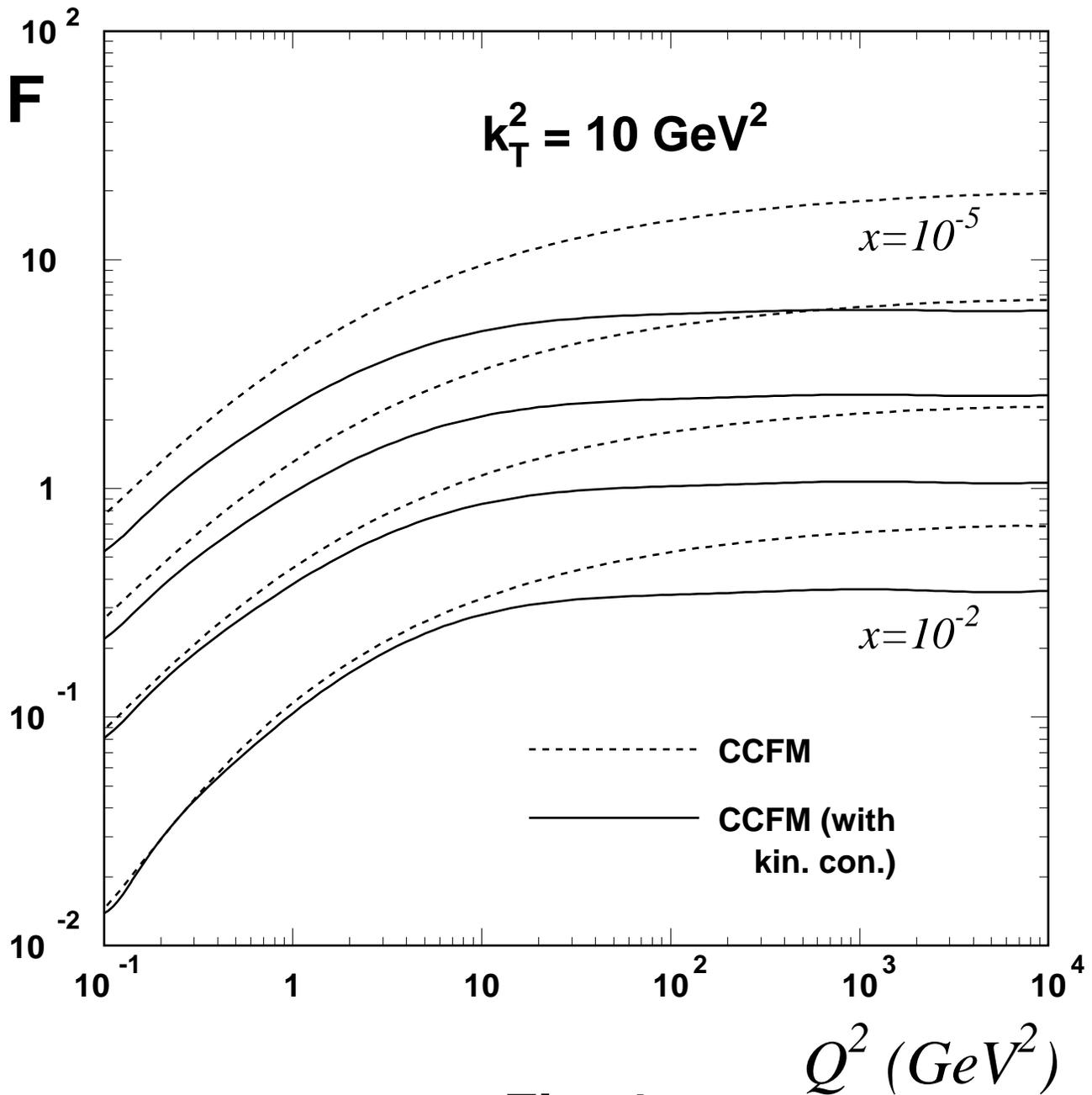
\item[Fig.\ 4] The $Q^2$ dependence of the gluon distribution $F
(x, k_T^2, Q^2)$ obtained by solving the CCFM equation with
(continuous curves) and without (dashed curves) the kinematic
constraint (\ref{a33}) included.  Results are shown for $x =
10^{-5}, 10^{-4}, 10^{-3}$ and $10^{-2}$.

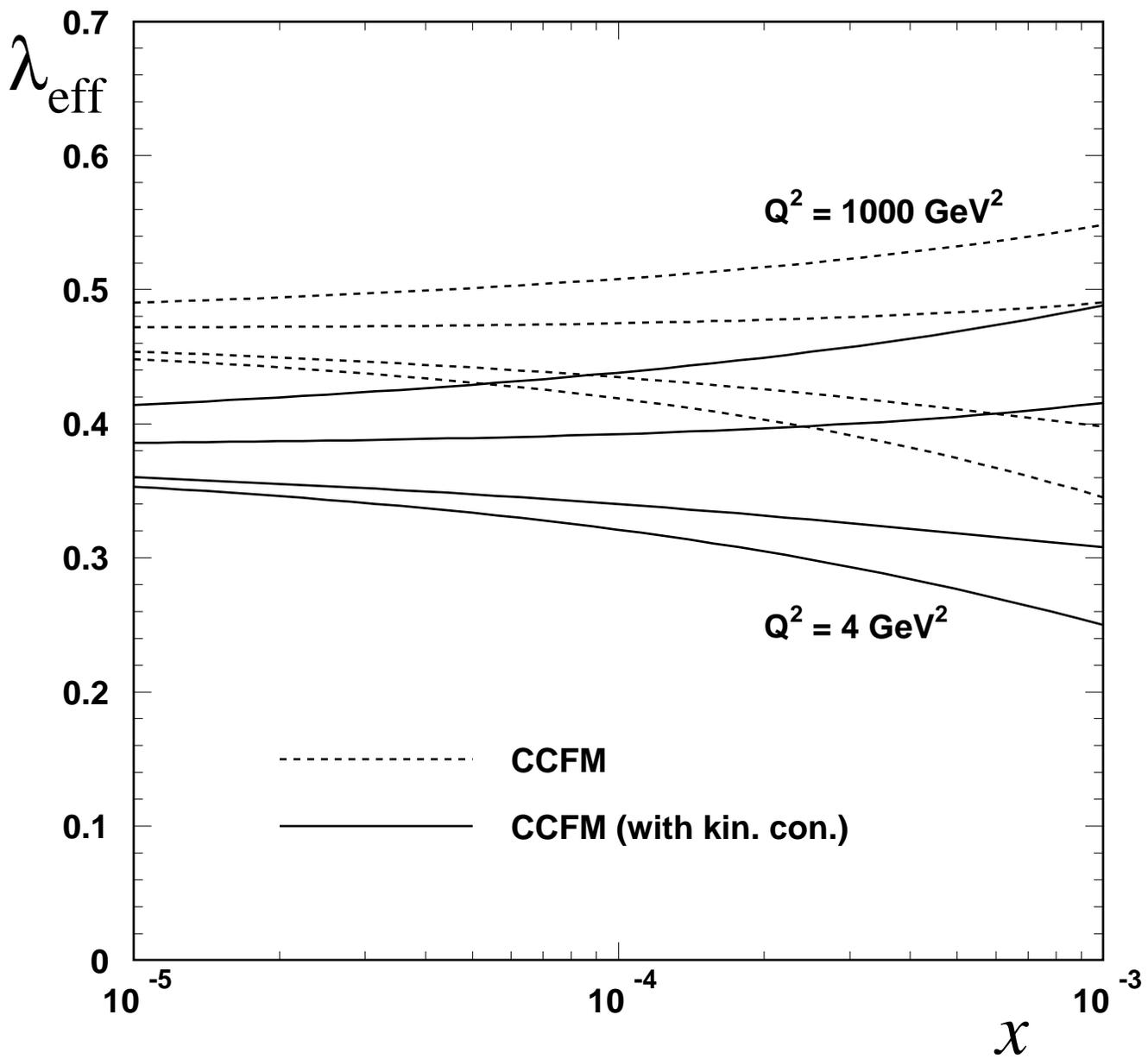
\item[Fig.\ 5] The effective exponent $\lambda_{\rm eff} \equiv
\partial \ln (xg)/\partial \ln (1/x)$ obtained by solving the
CCFM equation with (continuous curves) and without (dashed
curves) the kinematic constraint (\ref{a33}) included.  Results
are shown for $Q^2 = 1000, 100, 10$ and $4 \: {\rm GeV}^2$.

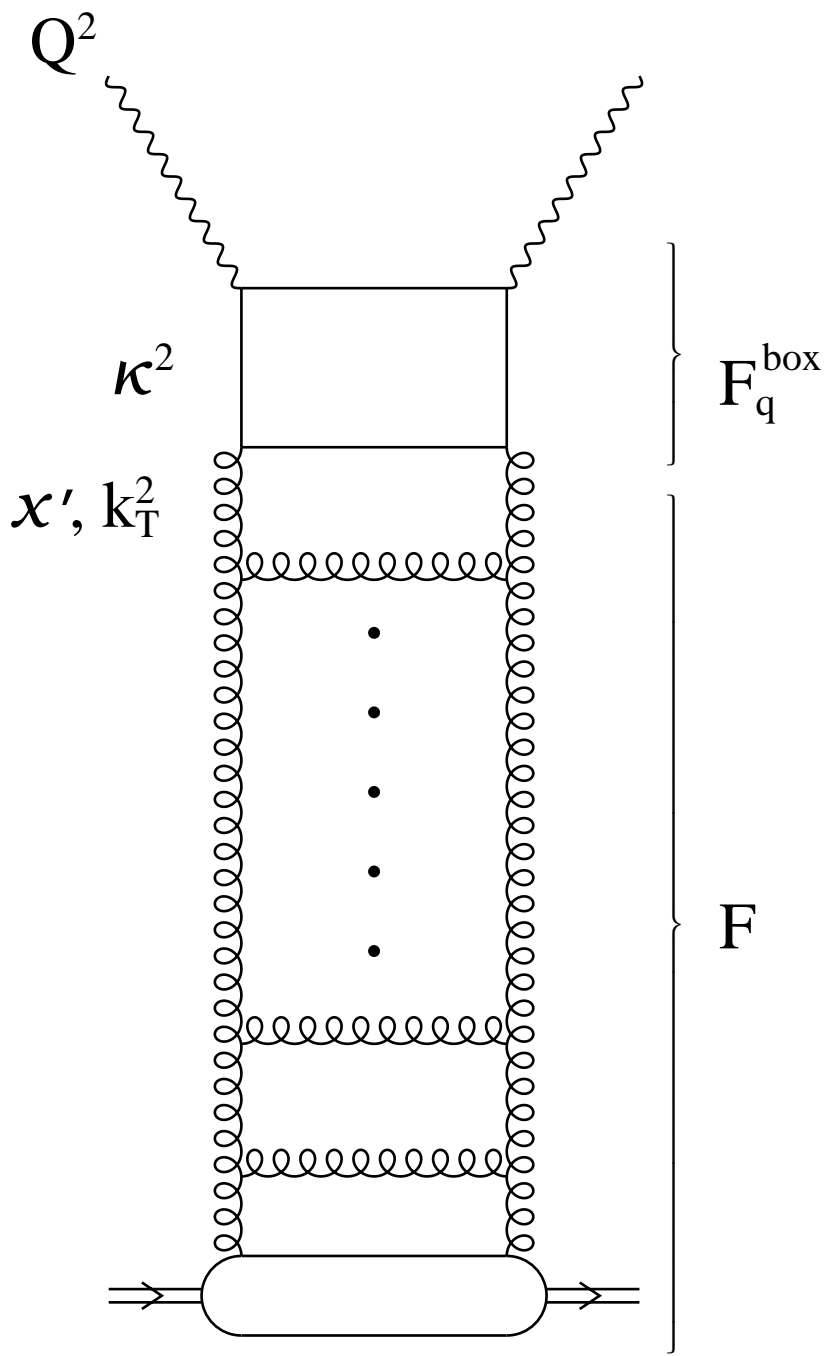
\item[Fig.\ 6] Pictorial representation of the $k_T$
factorization formula, that is of the convolution 
$ F_2 = \sum_q F
\otimes F_q^{\rm box}$ of (\ref{a40}).  $F (x^\prime, k_T^2,
\kappa^2)$ is the unintegrated gluon distribution and 
$ \sum_q F_q^{\rm box}$ 
is the off-shell gluon structure function, which at lowest
order is determined by the quark box (and \lq\lq crossed box")
contributions.

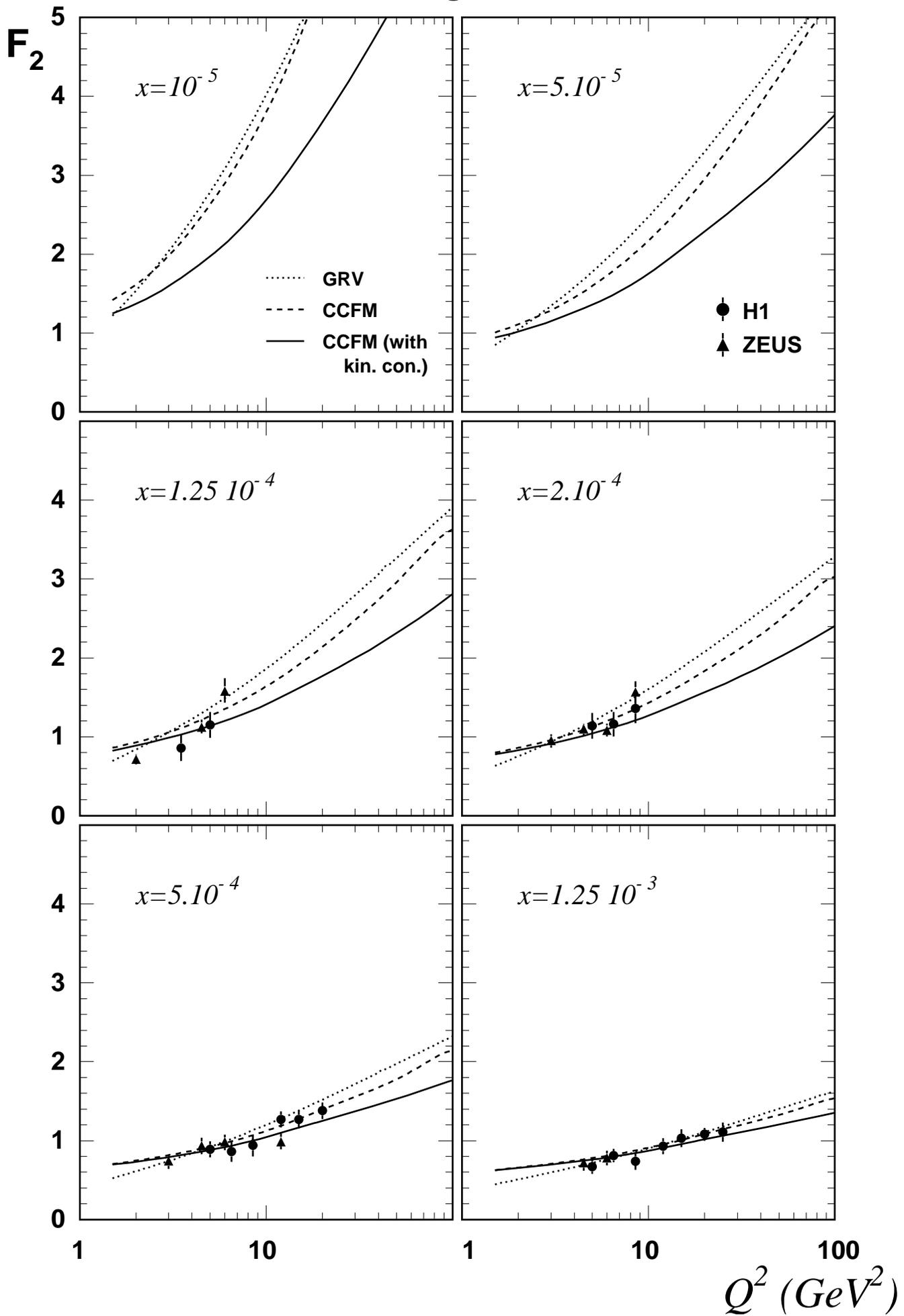
\item[Fig.\ 7] Predictions of the proton structure function $F_2
(x, Q^2)$ as a function of $\ln Q^2$, at fixed values of $x$,
compared with recent measurements made by the experiments at HERA
\cite{h1,zeus}.  The continuous and dashed curves are
respectively the predictions obtained from the CCFM equation, via
the $k_T$-factorization theorem, with and without the kinematic
constraint (\ref{a33}) incorporated.  The dotted curves are the
predictions obtained from the GRV set of partons \cite{grv}.

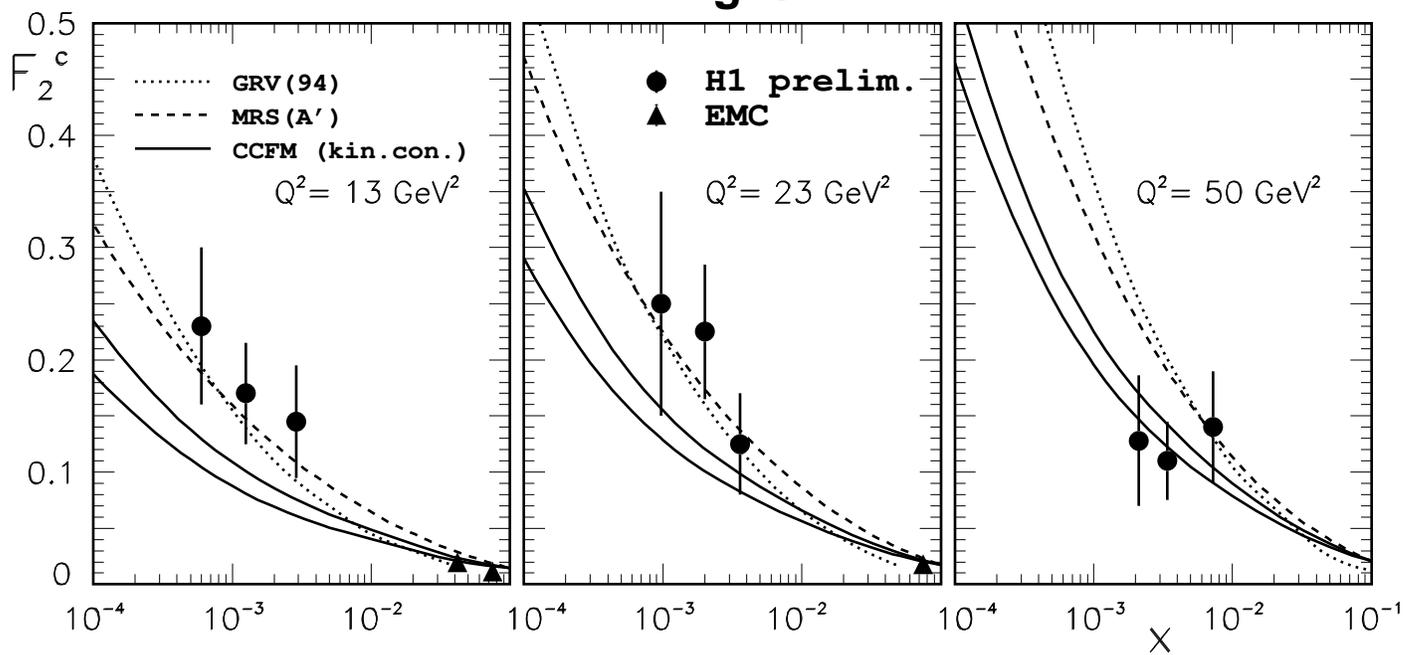
\item[Fig.\ 8] Predictions for the charm component $F_2^c$
of the proton structure function, $F_2$, compared to
recent preliminary H1 measurements \cite{adr} and older EMC data
\cite{emc}. 
The predictions were obtained by solving the
CCFM equation (with kinematic constraint) 
for the unintegrated gluon and then using the 
$k_T$-factorization formula with $m_c=1.4$ GeV (upper continuous
curve) and $m_c=1.7$ GeV (lower continuous curve)
Also shown are the next-to-leading order 
predictions \cite{v,mrsc} based on 
GRV \cite{grv} and MRS \cite{mrsa} partons.

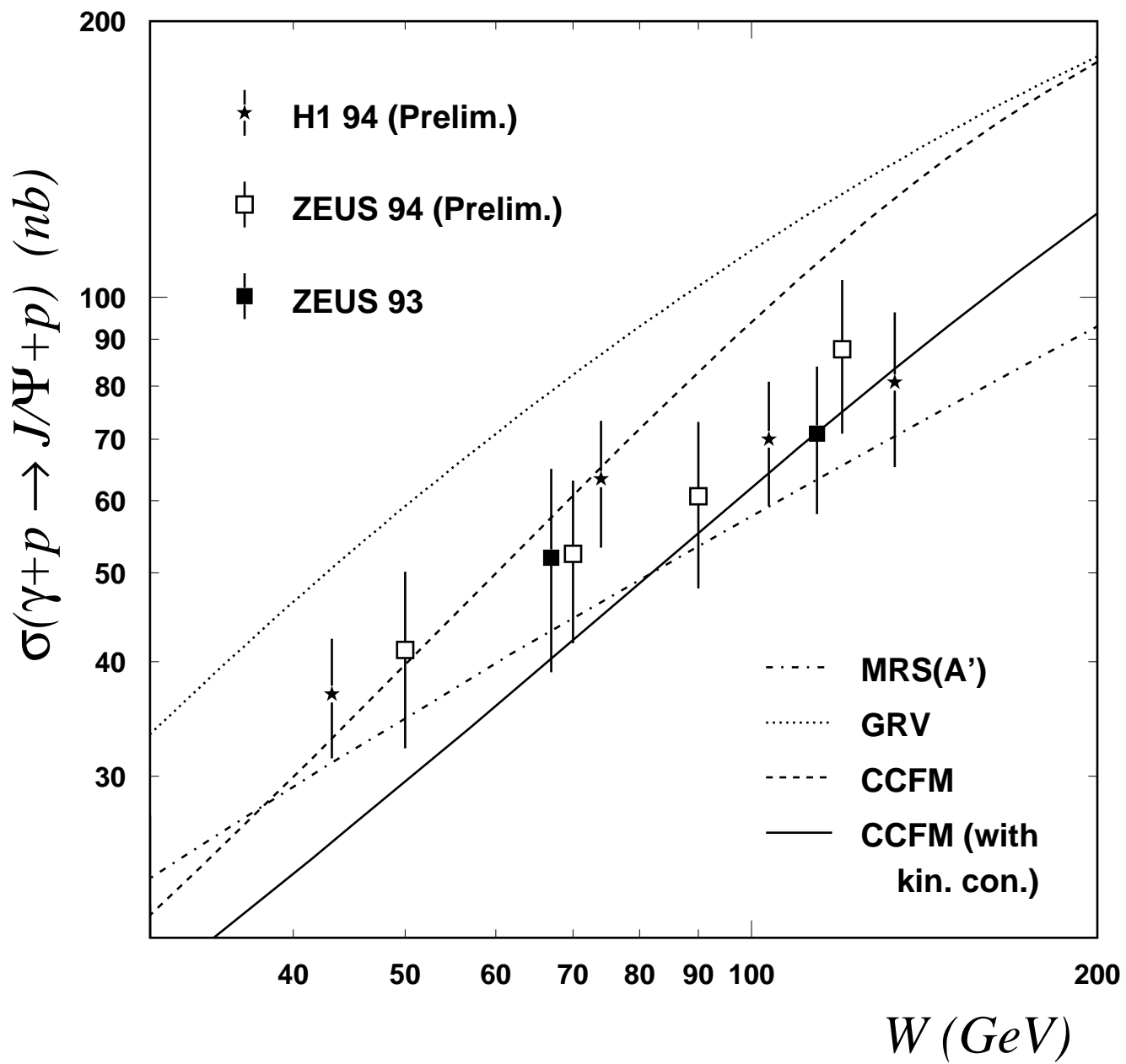
\item[Fig.\ 9] The measurements \cite{jpsi} of the cross section
for diffractive $J/\psi$ photoproduction compared with the
perturbative QCD description based on the unintegrated gluon
distribution obtained by solving the CCFM equation with
(continuous curve) and without (dashed curve) the kinematic
constraint (\ref{a33}) included.  The dotted and dash-dotted
curves are the predictions obtained from the GRV and
MRS(A$^\prime)$ set of partons \cite{grv,mrsa}, calculated as in
ref.\ \cite{rrml}.
\end{itemize}

\end{document}